\documentclass[twocolumn]{jpsj3}
\usepackage{txfonts}
\usepackage{bm}
\usepackage{color}
\usepackage{upgreek}

\begin{document}

\title{Spin-lattice-coupled helical magnetic order in breathing pyrochlore magnets, CuAlCr$_{4}$S$_{8}$ and CuGaCr$_{4}$S$_{8}$}

\author{Masaki Gen$^{1}$\thanks{masaki.gen@riken.jp}, Taro Nakajima$^{1,2}$, Hiraku Saito$^{2}$, Yusuke Tokunaga$^{3}$, and Taka-hisa Arima$^{1,3}$}
\inst{$^{1}$RIKEN Center for Emergent Matter Science (CEMS), Wako 351-0198, Japan\\
$^{2}$Institute for Solid State Physics, University of Tokyo, Kashiwa 277-8581, Japan\\
$^{3}$Department of Advanced Materials Science, University of Tokyo, Kashiwa 277-8561, Japan}

\abst{We report low-temperature powder X-ray and neutron diffraction studies on breathing pyrochlore magnets Cu{\it M}Cr$_{4}$S$_{8}$ ({\it M} = Al, Ga), which undergo a magnetic transition at $T_{\rm N} \approx 21$ and 31~K for {\it M} = Al and Ga, respectively.
X-ray diffraction reveals that the magnetic transition accompanies a structural transition from cubic $F{\overline 4}3m$ to polar orthorhombic $Imm2$ symmetry for both the compounds, with larger distortion observed for {\it M} = Ga at low temperatures.
Neutron diffraction reveals incommensurate magnetic modulation ${\mathbf Q} = (q_{\rm IC}, 0.5, 0)$ in the orthorhombic setting, where $q_{\rm IC} \approx 0.39$ and 0.31 for {\it M} = Al and Ga, respectively.
Our magnetic-structure analysis suggests cycloid-type magnetic order but not proper-screw type for both the compounds.
We find strong correlation between the local spin configuration and Cr--Cr bond lengths, indicating that the spin-lattice coupling as well as the magnetic frustration play an important role in determining the ground state.
Cu{\it M}Cr$_{4}$S$_{8}$ potentially offers a platform to explore magnetoelectric effects arising from the helimagnet driven electric polarity.}

\maketitle

\section{\label{Sec1}Introduction}
The breathing pyrochlore magnet, consisting of a three-dimensional network of alternately arranged small and large tetrahedra, has recently been recognized as a new class of frustrated spin systems \cite{2013_Oka}.
The introduction of inequivalent nearest-neighbor (NN) exchange interactions $J$ and $J'$ in the small and large tetrahedra, respectively, provides another degree of freedom absent in the regular pyrochlore lattice.
The emergence of unconventional ground states due to the breathing nature, such as the tetramer-singlet state \cite{2014_Kim}, commensurate hedgehog-antihedgehog lattice \cite{2021_Aoy_1}, and spin-lattice-coupled magnetic superstructures \cite{2021_Aoy_2} has been proposed.

{\it A}-site ordered chromium thiospinels with the Cr$^{3+}$ ion carrying an $S=3/2$ spin offer a fertile playground for systematically investigating the magnetism of the breathing pyrochlore classical Heisenberg models with various combinations of $J$ and $J'$ \cite{1970_Pin, 1976_Wil, 2018_Oka, 2018_Pok, 2019_Gho, 2020_Gen, 2021_Gao, 2022_Sha, 2022_Gen, 2023_Gen}.
Density functional theory (DFT) calculations predict that both $J$ and $J'$ are ferromagnetic (FM) in Li{\it M}Cr$_{4}$S$_{8}$ ({\it M} = Ga, In), whereas $J$ is antiferromagnetic (AFM) and $J'$ is FM in Cu{\it M}Cr$_{4}$S$_{8}$ ({\it M} = Ga, In) \cite{2019_Gho, 2023_Gen}.
If the FM $J'$ is dominant in the latter case, four spins in each large tetrahedron may behave as an effective $S=6$ spin \cite{2015_Ben}, which is observed as a FM cluster excitation in CuInCr$_{4}$S$_{8}$ \cite{2021_Gao}.
Relatively strong second-NN and third-NN AFM exchange interactions are also present in Cr thiospinels \cite{2019_Gho, 2023_Gen}, giving rise to the magnetic frustration pervading the entire lattice.

One important character of Cr spinels is a predominant role of the spin-lattice coupling due to the distance-sensitive direct exchange interaction between NN spins.
At low temperatures, Cr spinels undergo a magnetic transition accompanied by crystallographic symmetry breaking \cite{2008_Mar, 2018_Gao, 2009_Ji, 2005_Chu, 2006_Ued, 2007_Mat, 2016_Sah, 2015_Nil, 2009_Yok}.
Table~\ref{Tab_CrSpinel} summarizes the crystal structure and magnetic propagation vector in low-temperature phases of several related compounds, revealed by previous X-ray diffraction (XRD) and neutron diffraction (ND) experiments.
The change in the non-magnetic {\it A}-sites results in a variety of magnetic states with different crystallographic symmetry.
A phase separation is observed in some compounds \cite{2007_Mat, 2016_Sah, 2015_Nil, 2009_Yok}.
These observations suggest that the ground state of the Cr spinel is selected by a delicate balance between the exchange and elastic energy, and also significantly influenced by sample quality and strain.

In this study, we focus on the crystal and magnetic structures of the breathing pyrochlore magnets Cu{\it M}Cr$_{4}$S$_{8}$ ({\it M} = Al, Ga) at low temperatures.
Both compounds possess a cubic crystal structure with the $F{\overline 4}3m$ space group at room temperature \cite{2022_Sha, 2023_Gen}.
Upon cooling, a magnetic transition occurs at $T_{\rm N} \approx 21$~K for {\it M} = Al \cite{2022_Sha} and at $T_{\rm N} \approx 31$~K for {\it M} = Ga \cite{1970_Pin, 2023_Gen}, accompanied by a sudden drop in magnetic susceptibility.
The crystal and magnetic structures in the low-temperature phase have not been fully elucidated so far in spite of a few reports of X-ray and neutron studies \cite{1976_Wil, 2022_Sha, 2023_Gen}.
We reveal through powder XRD experiments that CuAlCr$_{4}$S$_{8}$ and CuGaCr$_{4}$S$_{8}$ commonly undergo a transition to an orthorhombic structure with the $Imm2$ space group at $T_{\rm N}$.
Furthermore, powder ND experiments reveal the emergence of cycloidal order with an incommensurate propagation vector of ${\mathbf Q} = (q_{\rm IC}, 0.5, 0)_{\rm o}$ in the orthorhombic setting, where $q_{\rm IC} \approx 0.39$ and 0.31 for {\it M} = Al and Ga, respectively.
The identified ground states are unique to the present compounds among the Cr spinel families (see Table~\ref{Tab_CrSpinel}), which can be attributed to the breathing degree of freedom in addition to the magnetic frustration and spin-lattice coupling. 

\begin{table*}[t]
\centering
\renewcommand{\arraystretch}{1.2}
\caption{Crystal structure and magnetic propagation vector of a magnetic ordering phase in various chromium spinel compounds. For LiGaCr$_{4}$O$_{8}$ and LiGaCr$_{4}$O$_{8}$, a phase separation of tetragonal and cubic phases was reported. The magnetic propagation vectors are based on the original cubic cell. ``+" indicates a superposition of multiple modulations, whereas ``\&'' indicates a phase separation.} 
\begin{tabular}{ccccc} \hline
~~ & ~Crystal structure~ & ~Magnetic propagation vector~ & ~Temperature~ & ~Reference \\ \hline
MgCr$_{2}$O$_{4}$ & ~Tetragonal $I4_{1}/amd$~($c/a = 0.9979$)~ & ~${\mathbf Q}_{1} = (0.5, 0.5, 0)$ + ${\mathbf Q}_{2} = (1, 0, 0.5)$~ & ~$T < T_{\rm N} = 12.4$~K~ & ~Refs.~\citen{2008_Mar, 2018_Gao} \\ \hline
ZnCr$_{2}$O$_{4}$ & ~Tetragonal $I{\overline 4}m2$~($c/a = 0.9982$)~ & ~${\mathbf Q}_{1} = (0.5, 0.5, 0)$ \& ${\mathbf Q}_{2} = (1, 0, 0.5)$~  & ~$T < T_{\rm N} = 12.5$~K~ & ~Ref.~\citen{ 2009_Ji} \\ \hline
CdCr$_{2}$O$_{4}$ & ~Tetragonal $I4_{1}/amd$~($c/a = 1.0043$)~ & ~${\mathbf Q} = (0, 0.09, 1)$~ & ~$T < T_{\rm N} = 7.8$~K~ & ~Ref.~\citen{2005_Chu} \\ \hline
HgCr$_{2}$O$_{4}$ & ~Orthorhombic $Fddd$~ & ~${\mathbf Q}_{1} = (0.5, 0, 1)$ \& ${\mathbf Q}_{2} = (1, 0, 0)$~ & ~$T < T_{\rm N} = 5.8$~K~ & ~Refs.~\citen{2006_Ued, 2007_Mat} \\ \hline
LiGaCr$_{4}$O$_{8}$ & \parbox{5.5cm}{\centering \vspace{+0.05cm}Tetragonal $I{\overline 4}m2$~($c/a = 1.0056$) \\ \vspace{+0.05cm}\& Cubic $F{\overline 4}3m$ \vspace{+0.1cm}} & \parbox{4cm}{\centering \vspace{+0.1cm} ${\mathbf Q}_{\rm tetra} = (1, 0, 0.5)$ \\ \vspace{+0.05cm}\& ${\mathbf Q}_{\rm cubic} = (1, 0, 0)$ \vspace{+0.1cm}}~ & ~$T < T_{\rm N} = 13 \sim 16$~K~ & ~Ref.~\citen{2016_Sah} \\ \hline
LiInCr$_{4}$O$_{8}$~ & \parbox{5.5cm}{\centering \vspace{+0.05cm}Tetragonal $I{\overline 4}m2$~($c/a = 0.9910$) \\ \vspace{+0.05cm}\& Cubic $F{\overline 4}3m$ \vspace{+0.1cm}} & \parbox{4cm}{\centering \vspace{+0.1cm} ${\mathbf Q}_{\rm tetra} = (1, 0, 0.2)$ \\ \vspace{+0.05cm}\& ${\mathbf Q}_{\rm cubic} = (1, 0, 0)$ \vspace{+0.1cm}}~ & ~$T < T_{\rm N} = 14.5$~K~ & ~Ref.~\citen{2015_Nil} \\ \hline
ZnCr$_{2}$S$_{4}$ & \parbox{5.7cm}{\centering \vspace{+0.1cm} Tetragonal $I4_{1}/amd$~($c/a = 0.9985$) \\ \vspace{+0.05cm}Orthorhombic $Imma$~($c/a = 1.0003, \gamma = 89.832^{\circ}$) \vspace{+0.1cm}} & \parbox{4.2cm}{\centering \vspace{+0.1cm} ${\mathbf Q}_{1} = (0, 0, 0.79)$ \& ${\mathbf Q}_{2} = (0.5, 0.5, 0)$ \\ \vspace{+0.05cm}${\mathbf Q} = (1, 0.5, 0)$ \vspace{+0.1cm}} & \parbox{2.7cm}{\centering \vspace{+0.1cm} $T_{\rm N2} < T < T_{\rm N1} = 15$~K \\ \vspace{+0.05cm}$T < T_{\rm N2} = 8$~K \vspace{+0.1cm}} & ~Ref.~\citen{2009_Yok} \\ \hline
CuAlCr$_{4}$S$_{8}$ &~Orthorhombic $Imm2$~($c/a = 1.0004, \gamma = 89.833^{\circ}$)~ & ~${\mathbf Q} = (0.89, 0.11, 0)$ [${\mathbf Q} = (0.39, 0.5, 0)_{\rm o}$]~ & ~$T < T_{\rm N} = 21$~K~ & ~This work \\ \hline
CuGaCr$_{4}$S$_{8}$ &~Orthorhombic $Imm2$~($c/a = 1.0005, \gamma = 89.743^{\circ}$)~ & ~${\mathbf Q} = (0.81, 0.19, 0)$ [${\mathbf Q} = (0.31, 0.5, 0)_{\rm o}$]~ & ~$T < T_{\rm N} = 31$~K~ & ~This work \\ \hline
\end{tabular}
\vspace{-0.5cm}
\label{Tab_CrSpinel}
\end{table*}

\begin{figure}[b]
\centering
\includegraphics[width=0.9\linewidth]{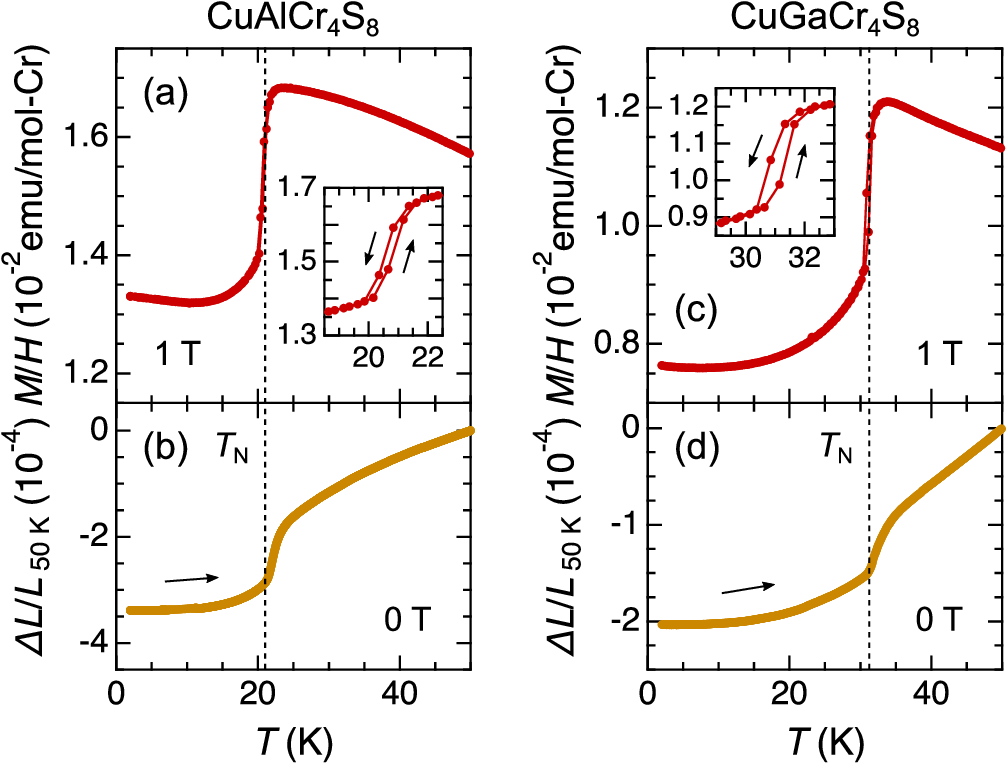}
\caption{Temperature dependence of magnetic susceptibility $M/H$ at 1~T (top) and thermal expansion $\Delta L/L$ at 0~T (bottom) for CuAlCr$_{4}$S$_{8}$ (left) and CuGaCr$_{4}$S$_{8}$ (right). The data of CuGaCr$_{4}$S$_{8}$ are excerpt from Ref.~\citen{2023_Gen}. Magnetization was measured during a cooling process down to 2 K and a subsequent warming process.
The inset in (a) and (c) displays an enlarged view of the $M/H$--$T$ curve in the vicinity of the magnetic transition temperature $T_{\rm N}$.}
\label{Fig1}
\end{figure}

\section{\label{Sec2}Method}
 
Polycrystalline samples of CuAlCr$_{4}$S$_{8}$ and CuGaCr$_{4}$S$_{8}$ were synthesized by the conventional solid-state reaction method.
A stoichiometric mixture of copper (99.99\%), aluminum (99.99\%)/gallium (99.999\%), chromium (99.99\%), and sulfur (99.99\%) powders was sealed in an evacuated quartz tube and heated at 900$^{\circ}$C for 48~h.
Subsequently, the sintering process was repeated twice at 1000$^{\circ}$C for 96~h.
Magnetization was measured using a superconducting quantum interference device (MPMS, Quantum Design).
Thermal expansion was measured by the fiber-Bragg-grating (FBG) method using an optical sensing instrument (Hyperion si155, LUNA) in a cryostat equipped with a superconducting magnet (Spectromag, Oxford).
The optical fiber was glued to the sintered samples using Stycast1266.

Powder XRD experiments were performed using a Rigaku SmartLab diffractometer at Materials Characterization Support Team, RIKEN CEMS.
The incident X-ray beam was monochromatized to Cu-$K\alpha_{1}$ radiation by a Johansson-type monochromator with a Ge(111) crystal.
The Rietveld analysis was performed using the RIETAN software \cite{RIETAN}.
Crystal structures were visualized by using the VESTA software \cite{VESTA}.
Powder ND experiments were performed at a triple-axis spectrometer PONTA(5G) installed in JRR-3 of the Japan Atomic Energy Agency.
The spectrometer was operated in the two-axis mode, and the horizontal collimation was 40'-40'-40'-40'.
An incident neutron beam with the wavelength of $2.392~\AA$ was obtained by a PG (002) monochromator.

\section{\label{Sec3}Results}

\subsection{\label{Sec3-1}Magnetization and thermal expansion}

Figure~\ref{Fig1} shows the temperature dependence of magnetic susceptibility $M/H$ measured at 1~T and thermal expansion $\Delta L/L$ at 0~T for Cu{\it M}Cr$_{4}$S$_{8}$ ({\it M} = Al, Ga).
The $M/H$--$T$ curve of CuAlCr$_{4}$S$_{8}$ well reproduces the data reported in Ref.~\citen{2022_Sha}.
For both the compounds, $M/H$ suddenly drops at $T_{\rm N}$, suggesting the development of AFM long-range order in a low-temperature phase.
This phase transition accompanies a hysteresis [see insets of Figs.~\ref{Fig1}(a) and \ref{Fig1}(c)], indicative of a first-order transition.
A steep decrease in the sample length can be seen above $T_{\rm N}$ in the $\Delta L/L$--$T$ curves, suggesting the occurrence of a structural transition associated with volume reduction.
As shown below, similar behaviors in $M/H$ and $\Delta L/L$ between CuAlCr$_{4}$S$_{8}$ and CuGaCr$_{4}$S$_{8}$ can be ascribed to the appearance of similar spin-lattice-coupled ground states. 

\begin{figure}[t]
\centering
\includegraphics[width=0.9\linewidth]{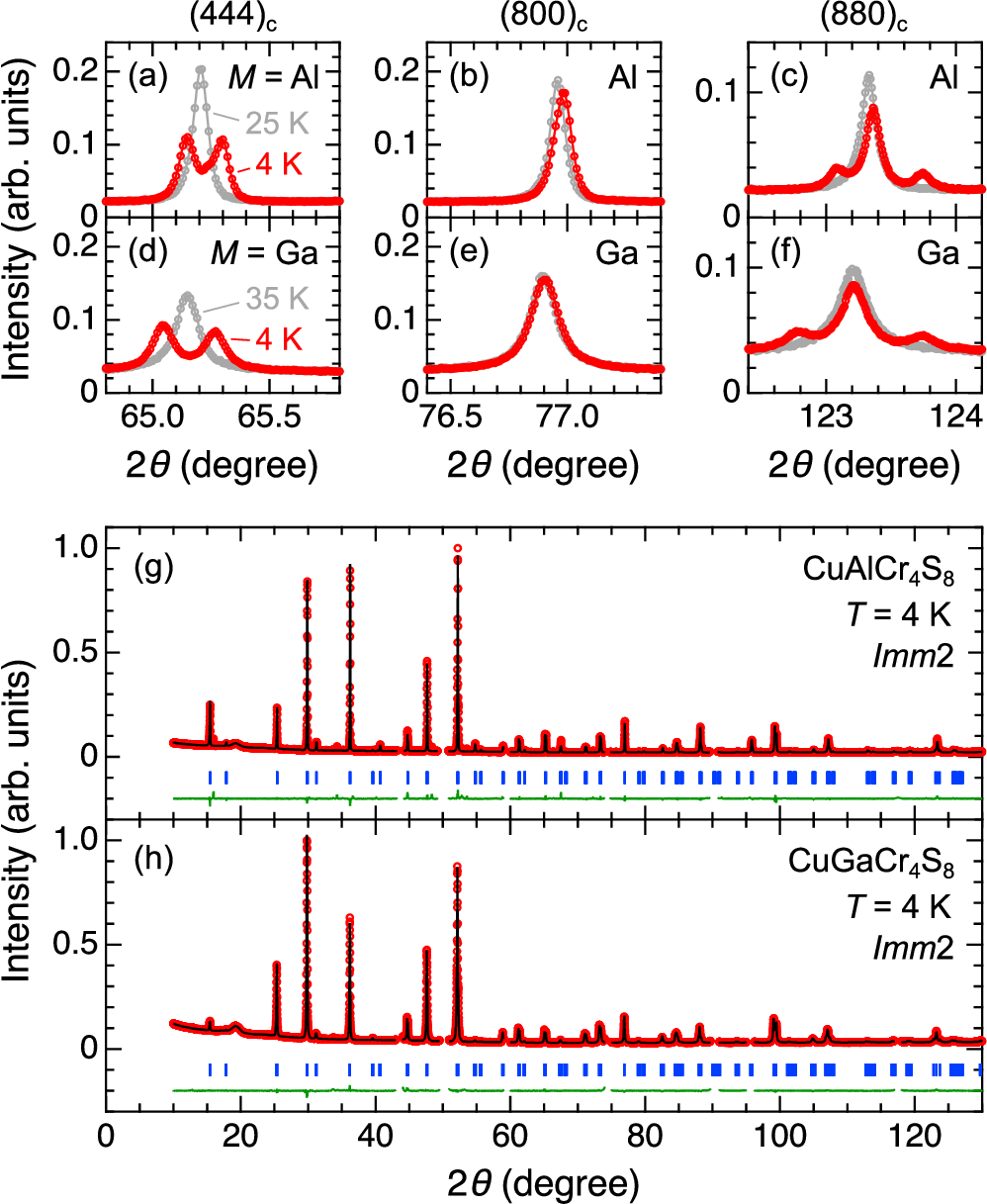}
\caption{Results of powder XRD experiments on Cu{\it M}Cr$_{4}$S$_{8}$ ({\it M} = Al, Ga) at low temperatures. [(a)--(f)] Experimental XRD profiles at selected Bragg positions at 25~K/35~K (gray) and 4~K (red) for [(a)--(c)] {\it M} = Al and [(d)--(f)] {\it M} = Ga, respectively. The indices, 800, 444, and 880, are defined based on the original cubic cell. The observed peak splittings are compatible with the orthorhombic $Imm2$ space group (No.~44). [(g)(h)] Experimental XRD patterns at 4~K (red open circles) and the calculated ones obtained by the Rietveld analysis assuming the $Imm2$ space group (black lines) for (g) {\it M} = Al and (h) {\it M} = Ga. Blue vertical bars indicate positions of the Bragg reflections, and the green line is a difference between the experimental and calculated patterns. Bragg peaks originating from the copper plate are removed from the analysis.}
\vspace{-0.5cm}
\label{Fig2}
\end{figure}

\subsection{\label{Sec3-2}Powder X-ray diffraction experiments}

\begin{figure*}[t]
\centering
\includegraphics[width=0.93\linewidth]{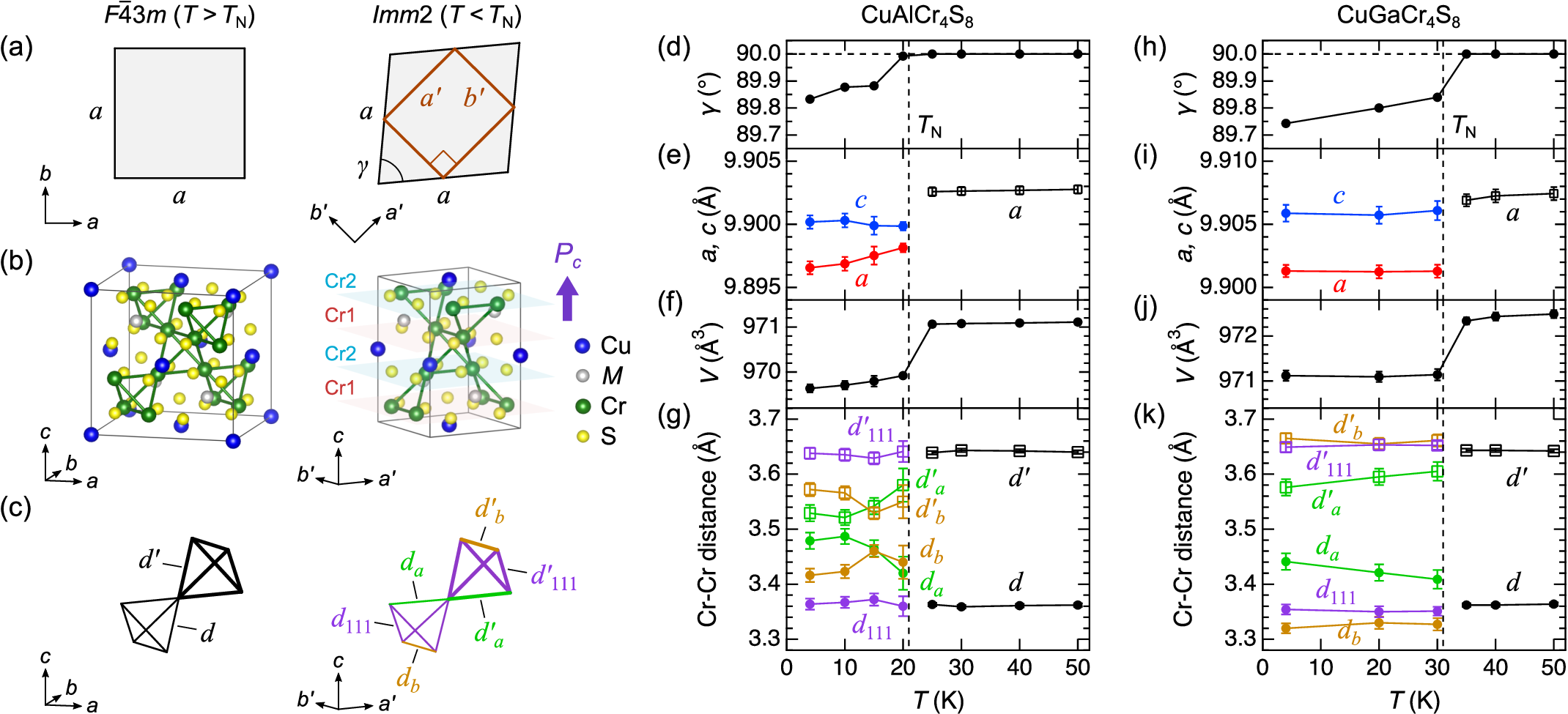}
\caption{Summary of crystallographic properties in Cu{\it M}Cr$_{4}$S$_{8}$ ({\it M} = Al, Ga). (a) Schematic representation of the lattice distortion in the $ab$ plane across $T_{\rm N}$. (b) Crystal structures within a unit cell for the $F{\overline 4}3m$ (left) and $Imm2$ (right) space groups. For the latter, the crystallographic polarity ($P_{c}$) exists along the $c$ axis. (c) Cr tetrahedral network and the definition of the NN Cr--Cr bond lengths in the $F{\overline 4}3m$ (left) and $Imm2$ (right) structures. Bonds of the same length are drawn in the same color and thickness. [(d)--(k)] Temperature dependence of (d)(h) lattice constants $\gamma$, (e)(i) $a$ and $c$, and (f)(j) volume $V$ of the original cubic cell, and (g)(k) the NN Cr--Cr bond lengths for {\it M} = Al (left) and {\it M} = Ga (right).}
\vspace{-0.5cm}
\label{Fig3}
\end{figure*}

Figures~\ref{Fig2}(a)--\ref{Fig2}(c) and \ref{Fig2}(d)--\ref{Fig2}(f) show changes in XRD peak profiles of selected Bragg reflections across $T_{\rm N}$ for CuAlCr$_{4}$S$_{8}$ and CuGaCr$_{4}$S$_{8}$, respectively.
Here, the indices are based on the original cubic cell.
We observe a common feature in the peak splitting patterns for the two compounds: (i) the $hhh$ reflection splits into two peaks with an intensity ratio of approximately 1:1, (ii) the $hh0$ reflection splits into three peaks with an intensity ratio of approximately 1:4:1, and (iii) the $h00$ reflection does not exhibit clear splitting.
These observations are compatible with rhombic distortion characterized by $a = b \neq c$, $\alpha = \beta = 90^{\circ}$, $\gamma \neq 90^{\circ}$ from the original cubic $F$ lattice, resulting in the orthorhombic $I$ lattice with a halved unit cell, as depicted in Fig.~\ref{Fig3}(a).
Since no additional Bragg peaks appear, we identify the crystallographic space group below $T_{\rm N}$ as $Imm2$ (No.~44).
The splitting widths of the $hhh$ and $hh0$ peaks, which reflect the deviation of the angle $\gamma$ from $90^{\circ}$, are larger for CuGaCr$_{4}$S$_{8}$ than for CuAlCr$_{4}$S$_{8}$ [Figs.~\ref{Fig2}(a), \ref{Fig2}(c), \ref{Fig2}(d), and \ref{Fig2}(f)].
This indicates that CuGaCr$_{4}$S$_{8}$ exhibits larger rhombic distortion in the $ab$ plane compared to CuAlCr$_{4}$S$_{8}$.
A structural transition from $F{\overline 4}3m$ to $Imm2$ symmetry has been reported in lacunar spinels GeV$_{4}$S$_{8}$ \cite{2008_Bic} and GaMo$_{4}$Se$_{8}$ \cite{2020_Sch}, where the orbital degree of freedom plays an important role.
On the other hand, only the interplay between spin and lattice degrees of freedom should be a driving force of the structural transition in Cu{\it M}Cr$_{4}$S$_{8}$.

\begin{table}[t]
\centering
\renewcommand{\arraystretch}{1.2}
\caption{Structural parameters of CuAlCr$_{4}$S$_{8}$ obtained from the Rietveld analysis at 25~K (top) and 4~K (bottom) based on the $F{\overline 4}3m$ (No.~216) and $Imm2$ space group (No.~44), respectively. The lattice constants are $a = 9.90259(35)~\AA$ at 25~K, and $a' = 7.00809(37)~\AA$, $b' = 6.98774(37)~\AA$, $c = 9.90018(53)~\AA$ at 4~K.}
\begin{tabular}{ccccccc} \hline
\multicolumn{7}{l}{\hspace{-0.1cm}$T = 25$~K ($R_{\rm wp} = 4.82$, $R_{\rm p} = 2.77$, $R_{\rm e} = 2.60$, $S = 1.85$)}\\
\hspace{-0.1cm}Atom & \hspace{-0.3cm}Site & \hspace{-0.3cm}Occ. & \hspace{-0.2cm}$x$ & \hspace{-0.2cm}$y$ & \hspace{-0.2cm}$z$ & \hspace{-0.2cm}$B$~(\AA$^{2}$)\hspace{-0.1cm} \\ \hline
\hspace{-0.1cm}Cu & \hspace{-0.3cm}$4a$ & \hspace{-0.3cm}1 & \hspace{-0.2cm}0 & \hspace{-0.2cm}0 & \hspace{-0.2cm}0 & \hspace{-0.2cm}0.380(54)\hspace{-0.1cm} \\
\hspace{-0.1cm}Al & \hspace{-0.3cm}$4d$ & \hspace{-0.3cm}1 & \hspace{-0.2cm}0.75 & \hspace{-0.2cm}0.75 & \hspace{-0.2cm}0.75 & \hspace{-0.2cm}0.015(93)\hspace{-0.1cm} \\
\hspace{-0.1cm}Cr & \hspace{-0.3cm}$16e$ & \hspace{-0.3cm}1 & \hspace{-0.2cm}0.37008(12) & \hspace{-0.2cm}$x$ & \hspace{-0.2cm}$x$ & \hspace{-0.2cm}0.391(19)\hspace{-0.1cm} \\
\hspace{-0.1cm}S1 & \hspace{-0.3cm}$16e$ & \hspace{-0.3cm}1 & \hspace{-0.2cm}0.13303(13) & \hspace{-0.2cm}$x$ & \hspace{-0.2cm}$x$ & \hspace{-0.2cm}0.2 (fix)\hspace{-0.1cm} \\
\hspace{-0.1cm}S2 & \hspace{-0.3cm}$16e$ & \hspace{-0.3cm}1 & \hspace{-0.2cm}0.61671(10) & \hspace{-0.2cm}$x$ & \hspace{-0.2cm}$x$ & \hspace{-0.2cm}0.2 (fix)\hspace{-0.1cm} \\ \hline
\multicolumn{7}{l}{\hspace{-0.1cm}$T = 4$~K ($R_{\rm wp} = 5.13$, $R_{\rm p} = 2.99$, $R_{\rm e} = 2.61$, $S = 1.97$)}\\
\hspace{-0.1cm}Atom & \hspace{-0.3cm}Site & \hspace{-0.3cm}Occ. & \hspace{-0.2cm}$x$ & \hspace{-0.2cm}$y$ & \hspace{-0.2cm}$z$ & \hspace{-0.2cm}$B$~(\AA$^{2}$)\hspace{-0.1cm} \\ \hline
\hspace{-0.1cm}Cu & \hspace{-0.3cm}$2a$ & \hspace{-0.3cm}1 & \hspace{-0.2cm}0.5 & \hspace{-0.2cm}0.5 & \hspace{-0.2cm}0 & \hspace{-0.2cm}0.426(66)\hspace{-0.1cm} \\
\hspace{-0.1cm}Al & \hspace{-0.3cm}$2b$ & \hspace{-0.3cm}1 & \hspace{-0.2cm}0 & \hspace{-0.2cm}0.5 & \hspace{-0.2cm}0.75 & \hspace{-0.2cm}0.063(118)\hspace{-0.1cm} \\
\hspace{-0.1cm}Cr1 & \hspace{-0.3cm}$4c$ & \hspace{-0.3cm}1 & \hspace{-0.2cm}0.5 & \hspace{-0.2cm}0.24442(83) & \hspace{-0.2cm}0.63480(80) & \hspace{-0.2cm}0.321(60)\hspace{-0.1cm} \\
\hspace{-0.1cm}Cr2 & \hspace{-0.3cm}$4d$ & \hspace{-0.3cm}1 & \hspace{-0.2cm}0.25181(100) & \hspace{-0.2cm}0 & \hspace{-0.2cm}0.86890(89) & \hspace{-0.2cm}0.174(55)\hspace{-0.1cm} \\
\hspace{-0.1cm}S1 & \hspace{-0.3cm}$4c$ & \hspace{-0.3cm}1 & \hspace{-0.2cm}0 & \hspace{-0.2cm}0.27483(85) & \hspace{-0.2cm}0.36926(90) & \hspace{-0.2cm}0.2 (fix)\hspace{-0.1cm} \\
\hspace{-0.1cm}S2 & \hspace{-0.3cm}$4c$ & \hspace{-0.3cm}1 & \hspace{-0.2cm}0 & \hspace{-0.2cm}0.24084(79) & \hspace{-0.2cm}0.88640(77) & \hspace{-0.2cm}0.2 (fix)\hspace{-0.1cm} \\
\hspace{-0.1cm}S3 & \hspace{-0.3cm}$4d$ & \hspace{-0.3cm}1 & \hspace{-0.2cm}0.23999(98) & \hspace{-0.2cm}0 & \hspace{-0.2cm}0.11376(77) & \hspace{-0.2cm}0.2 (fix)\hspace{-0.1cm} \\
\hspace{-0.1cm}S4 & \hspace{-0.3cm}$4d$ & \hspace{-0.3cm}1 & \hspace{-0.2cm}0.26817(105) & \hspace{-0.2cm}0 & \hspace{-0.2cm}0.63111(86) & \hspace{-0.2cm}0.2 (fix)\hspace{-0.1cm} \\ \hline
\end{tabular}
\vspace{-0.5cm}
\label{Tab_Al}
\end{table}

\begin{table}[t]
\centering
\renewcommand{\arraystretch}{1.2}
\caption{Structural parameters of CuGaCr$_{4}$S$_{8}$ obtained from the Rietveld analysis at 35~K (top) and 4~K (bottom) based on the $F{\overline 4}3m$ (No.~216) and $Imm2$ space group (No.~44), respectively. The lattice constants are $a = 9.90691(48)~\AA$ at 35~K, and $a' = 7.01696(48)~\AA$, $b' = 6.98556(47)~\AA$, $c = 9.90588(67)~\AA$ at 4~K.}
\begin{tabular}{ccccccc} \hline
\multicolumn{7}{l}{\hspace{-0.1cm}$T = 35$~K ($R_{\rm wp} = 3.36$, $R_{\rm p} = 2.24$, $R_{\rm e} = 2.90$, $S = 1.16$)}\\
\hspace{-0.1cm}Atom & \hspace{-0.3cm}Site & \hspace{-0.3cm}Occ. & \hspace{-0.2cm}$x$ & \hspace{-0.2cm}$y$ & \hspace{-0.2cm}$z$ & \hspace{-0.2cm}$B$~(\AA$^{2}$)\hspace{-0.1cm} \\ \hline
\hspace{-0.1cm}Cu & \hspace{-0.3cm}$4a$ & \hspace{-0.3cm}1 & \hspace{-0.2cm}0 & \hspace{-0.2cm}0 & \hspace{-0.2cm}0 & \hspace{-0.2cm}0.720(106)\hspace{-0.1cm} \\
\hspace{-0.1cm}Ga & \hspace{-0.3cm}$4d$ & \hspace{-0.3cm}1 & \hspace{-0.2cm}0.75 & \hspace{-0.2cm}0.75 & \hspace{-0.2cm}0.75 & \hspace{-0.2cm}0.087(70)\hspace{-0.1cm} \\
\hspace{-0.1cm}Cr & \hspace{-0.3cm}$16e$ & \hspace{-0.3cm}1 & \hspace{-0.2cm}0.36999(13) & \hspace{-0.2cm}$x$ & \hspace{-0.2cm}$x$ & \hspace{-0.2cm}0.214(17)\hspace{-0.1cm} \\
\hspace{-0.1cm}S1 & \hspace{-0.3cm}$16e$ & \hspace{-0.3cm}1 & \hspace{-0.2cm}0.13124(15) & \hspace{-0.2cm}$x$ & \hspace{-0.2cm}$x$ & \hspace{-0.2cm}0.2 (fix)\hspace{-0.1cm} \\
\hspace{-0.1cm}S2 & \hspace{-0.3cm}$16e$ & \hspace{-0.3cm}1 & \hspace{-0.2cm}0.61530(17) & \hspace{-0.2cm}$x$ & \hspace{-0.2cm}$x$ & \hspace{-0.2cm}0.2 (fix)\hspace{-0.1cm} \\ \hline
\multicolumn{7}{l}{\hspace{-0.1cm}$T = 4$~K ($R_{\rm wp} = 2.85$, $R_{\rm p} = 1.79$, $R_{\rm e} = 2.90$, $S = 0.97$)}\\
\hspace{-0.1cm}Atom & \hspace{-0.3cm}Site & \hspace{-0.3cm}Occ. & \hspace{-0.2cm}$x$ & \hspace{-0.2cm}$y$ & \hspace{-0.2cm}$z$ & \hspace{-0.2cm}$B$~(\AA$^{2}$)\hspace{-0.1cm} \\ \hline
\hspace{-0.1cm}Cu & \hspace{-0.3cm}$2a$ & \hspace{-0.3cm}1 & \hspace{-0.2cm}0.5 & \hspace{-0.2cm}0.5 & \hspace{-0.2cm}0 & \hspace{-0.2cm}0.157(102)\hspace{-0.1cm} \\
\hspace{-0.1cm}Ga & \hspace{-0.3cm}$2b$ & \hspace{-0.3cm}1 & \hspace{-0.2cm}0 & \hspace{-0.2cm}0.5 & \hspace{-0.2cm}0.75 & \hspace{-0.2cm}0.454(84)\hspace{-0.1cm} \\
\hspace{-0.1cm}Cr1 & \hspace{-0.3cm}$4c$ & \hspace{-0.3cm}1 & \hspace{-0.2cm}0.5 & \hspace{-0.2cm}0.23765(64) & \hspace{-0.2cm}0.63302(66) & \hspace{-0.2cm}0.040(44)\hspace{-0.1cm} \\
\hspace{-0.1cm}Cr2 & \hspace{-0.3cm}$4d$ & \hspace{-0.3cm}1 & \hspace{-0.2cm}0.25482(105) & \hspace{-0.2cm}0 & \hspace{-0.2cm}0.87053(88) & \hspace{-0.2cm}0.344(49)\hspace{-0.1cm} \\
\hspace{-0.1cm}S1 & \hspace{-0.3cm}$4c$ & \hspace{-0.3cm}1 & \hspace{-0.2cm}0 & \hspace{-0.2cm}0.26689(83) & \hspace{-0.2cm}0.37387(93) & \hspace{-0.2cm}0.2 (fix)\hspace{-0.1cm} \\
\hspace{-0.1cm}S2 & \hspace{-0.3cm}$4c$ & \hspace{-0.3cm}1 & \hspace{-0.2cm}0 & \hspace{-0.2cm}0.23284(71) & \hspace{-0.2cm}0.89064(93) & \hspace{-0.2cm}0.2 (fix)\hspace{-0.1cm} \\
\hspace{-0.1cm}S3 & \hspace{-0.3cm}$4d$ & \hspace{-0.3cm}1 & \hspace{-0.2cm}0.23080(107) & \hspace{-0.2cm}0 & \hspace{-0.2cm}0.11709(82) & \hspace{-0.2cm}0.2 (fix)\hspace{-0.1cm} \\
\hspace{-0.1cm}S4 & \hspace{-0.3cm}$4d$ & \hspace{-0.3cm}1 & \hspace{-0.2cm}0.26426(115) & \hspace{-0.2cm}0 & \hspace{-0.2cm}0.63118(95) & \hspace{-0.2cm}0.2 (fix)\hspace{-0.1cm} \\ \hline
\end{tabular}
\vspace{-0.5cm}
\label{Tab_Ga}
\end{table}

On the basis of the above-mentioned considerations, we perform a Rietveld analysis of the XRD patterns at low temperatures.
A satisfactory fit is achieved by assuming the $F{\overline 4}3m$ space group above $T_{\rm N}$ and the $Imm2$ space group below $T_{\rm N}$, as shown in Figs.~\ref{Fig2}(g) and \ref{Fig2}(h).
Tables~\ref{Tab_Al} and \ref{Tab_Ga} summarize the obtained structural parameters and reliability factors for CuAlCr$_{4}$S$_{8}$ and CuGaCr$_{4}$S$_{8}$, respectively.
As depicted in the right panel of Fig.~\ref{Fig3}(a), the lattice constants of the orthorhombic cell are defined such that $a' > b'$, where the $a'$ and $b'$ axes are set by rotating the cubic $a$ and $b$ axes by $45^{\circ}$.
The crystal structure within the corresponding unit cell for each space group is schematically illustrated in Fig.~\ref{Fig3}(b).
In the $Imm2$ structure, there exist two types of Cr sites, Cr1 and Cr2, alternately stacking along the $c$ axis.
Notably, the averaged $z$ coordinates of the Cr and S atoms deviate from $z = 0.5$ below $T_{\rm N}$, leading to a polar structure along the $c$ axis.
The rhombic distortion results in three types of NN Cr--Cr bonds within each tetrahedron, as shown in the right panel of Fig.~\ref{Fig3}(c).
In the following, we define the Cr--Cr bond lengths along the $a'$, $b'$, and four equivalent $\langle111\rangle$ axes in the orthorhombic setting below $T_{\rm N}$ as $d_{a}$ ($d'_{a}$), $d_{b}$ ($d'_{b}$), and $d_{111}$ ($d'_{111}$) in each small (large) tetrahedron.

Figures~\ref{Fig3}(d)--\ref{Fig3}(k) show the temperature dependence of various crystallographic parameters for CuAlCr$_{4}$S$_{8}$ and CuGaCr$_{4}$S$_{8}$.
The two compounds exhibit comparable volume reduction, $\Delta V/V \approx 1.2 \times 10^{-3}$, across $T_{\rm N}$ [Figs.~\ref{Fig3}(f) and \ref{Fig3}(j)].
Meanwhile, the lattice distortion is more significant in CuGaCr$_{4}$S$_{8}$ compared to CuAlCr$_{4}$S$_{8}$: $c/a = 1.0004$ and $\gamma = 89.833^{\circ}$ for CuAlCr$_{4}$S$_{8}$, and $c/a = 1.0005$ and $\gamma = 89.743^{\circ}$ for CuGaCr$_{4}$S$_{8}$ at 4~K [Figs.~\ref{Fig3}(d), \ref{Fig3}(e), \ref{Fig3}(h), and \ref{Fig3}(i)].
Above $T_{\rm N}$, the ratios of the two kinds of NN Cr--Cr bond lengths, $d$ and $d'$, are nearly identical between the two compounds: $d'/d = 1.082(3)$ for CuAlCr$_{4}$S$_{8}$ at 25~K and $d'/d = 1.084(3)$ for CuGaCr$_{4}$S$_{8}$ at 35~K.
Below $T_{\rm N}$, however, the difference between $d_{a}$ and $d'_{a}$, as well as between $d_{b}$ and $d'_{b}$, becomes much smaller for CuAlCr$_{4}$S$_{8}$ than for CuGaCr$_{4}$S$_{8}$ [Figs.~\ref{Fig3}(g) and \ref{Fig3}(k)].
In other words, the breathing of the pyrochlore network is greatly suppressed below $T_{\rm N}$ for CuAlCr$_{4}$S$_{8}$.
The resulting volume ratio of the larger tetrahedron to the smaller one below $T_{\rm N}$ is 1.21 for CuAlCr$_{4}$S$_{8}$ and 1.27 for CuGaCr$_{4}$S$_{8}$.

\subsection{\label{Sec3-3}Powder neutron diffraction experiments}

\begin{figure}[b]
\centering
\includegraphics[width=0.93\linewidth]{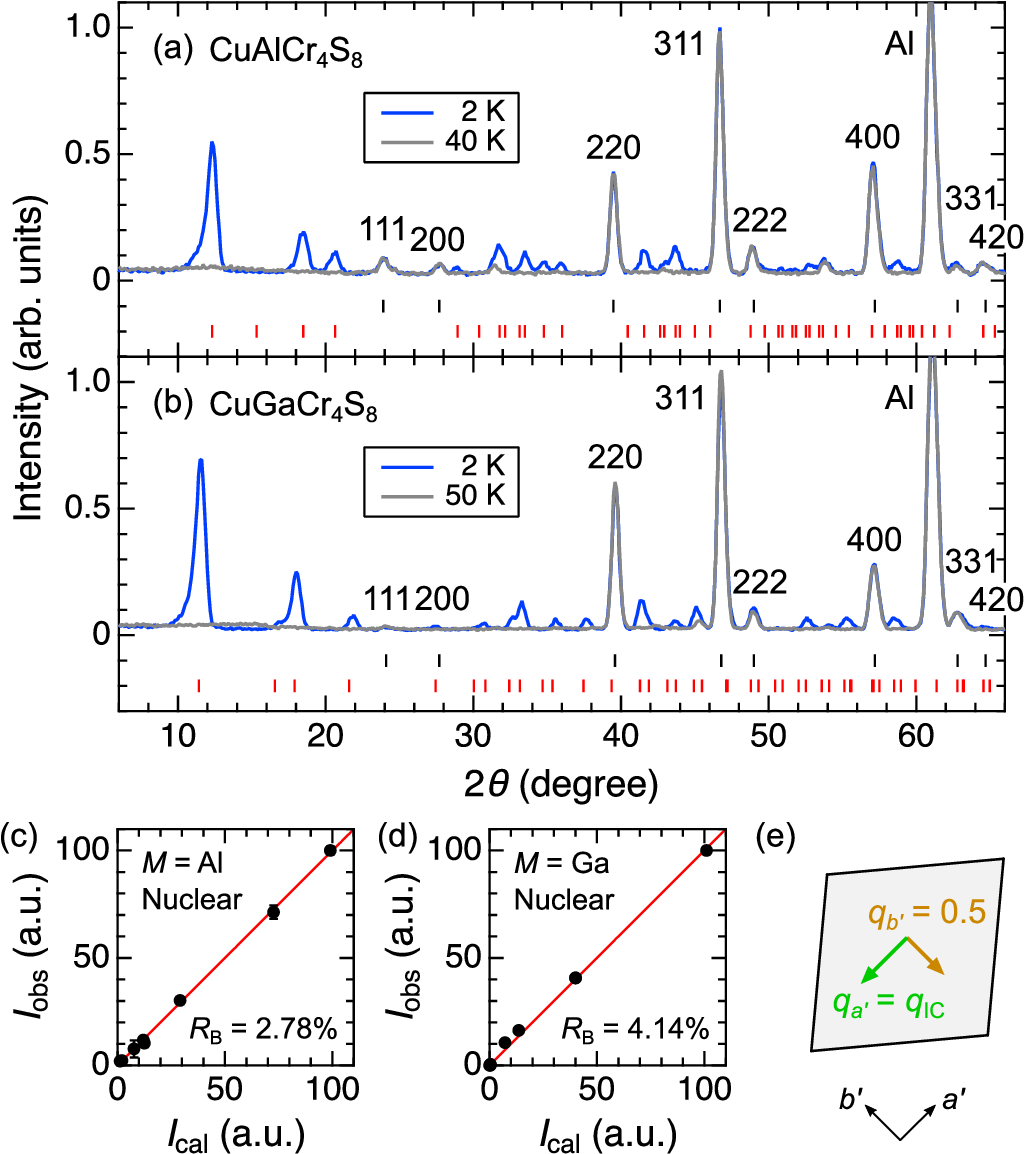}
\caption{[(a)(b)] Experimental powder ND patterns at 40~K/50~K (gray) and 2~K (blue) for (a) CuAlCr$_{4}$S$_{8}$ and (b) CuGaCr$_{4}$S$_{8}$. The strong peak at $2\theta = 61^{\circ}$ originates from aluminum. The indices of the nuclear Bragg peaks are defined in the cubic setting. For CuGaCr$_{4}$S$_{8}$, the 200 and 420 peaks are not observed due to the close scattering length of Cu and Ga atoms. Black and red vertical bars indicate positions of the nuclear and magnetic Bragg reflections, respectively. [(c)(d)] $I_{\rm obs}$ versus $I_{\rm cal}$ plot of the nuclear Bragg peaks at 2~K for (c) CuAlCr$_{4}$S$_{8}$ and (d) CuGaCr$_{4}$S$_{8}$. For evaluating $I_{\rm obs}$, Lorentz factor correction is performed by multiplying $\sin \theta \sin 2\theta$ to the observed intensity. $I_{\rm cal}$ is calculated using the structural parameters at 4~K (Tables~\ref{Tab_Al} and \ref{Tab_Ga}). (e) Orientation of the incommensurate ($q_{a'} = q_{\rm IC}$) and commensurate ($q_{b'} = 0.5$) magnetic modulation imposed on the orthorhombic cell.}
\label{Fig4}
\end{figure}

Figures~\ref{Fig4}(a) and \ref{Fig4}(b) show the experimental powder ND patterns for {\it M} = Al and Ga, respectively.
Here, the splitting is not observed for any nuclear Bragg peaks at 2~K due to the rather low $Q$ resolution of the instrument.
Figures.~\ref{Fig4}(c) and \ref{Fig4}(d) compare the observed intensities $I_{\rm obs}$ of nuclear Bragg peaks after the Lorentz factor correction with the calculated ones $I_{\rm cal}$ based on the 4-K crystal structure revealed by the XRD (Tables~\ref{Tab_Al} and \ref{Tab_Ga}).
A least-squares fit confirms good agreement between $I_{\rm obs}$ and $I_{\rm cal}$ for both the compounds; the intensity reliability factor $R_{\rm B}$ defined as $R_{\rm B} \equiv (\sum_{K} |I_{\rm obs}(K) - I_{\rm cal}(K)|)/\sum_{K} I_{\rm obs}(K)$, where the sums are taken over 8 nuclear reflections, is 2.78\% and 4.14\% for {\it M} = Al and Ga, respectively. 
The obtained scale factors are used for the subsequent magnetic-structure analysis.

As shown in Figs.~\ref{Fig4}(a) and \ref{Fig4}(b), we observe a number of additional Bragg reflections of magnetic origin at 2~K.
All of theses peaks can be indexed by a single magnetic propagation vector, ${\mathbf Q} \approx (0.89, 0.11, 0)_{\rm c}$ for CuAlCr$_{4}$S$_{8}$ and ${\mathbf Q} \approx (0.81, 0.19, 0)_{\rm c}$ for CuGaCr$_{4}$S$_{8}$ in the cubic setting.
The latter one is almost consistent with previously reported ${\mathbf Q} = (0.80, 0.18, 0)_{\rm c}$ for CuGaCr$_{4}$S$_{8}$ \cite{1976_Wil}.
Remarkably, the identified ${\mathbf Q}$ vectors can be represented as ${\mathbf Q} = (q_{\rm IC}, 0.5, 0)_{\rm o}$ in the orthorhombic setting, where $q_{\rm IC} \approx 0.39$ for CuAlCr$_{4}$S$_{8}$ and $q_{\rm IC} \approx 0.31$ for CuGaCr$_{4}$S$_{8}$.
Considering the exchange-striction mechanism where the AFM configuration favors a shorter Cr--Cr bond length due to the predominant direct exchange coupling, we presume that the longer-period incommensurate modulation ($q_{a'} = q_{\rm IC}$) propagates along the longer diagonal ($a'$ axis) and the shorter-period commensurate modulation ($q_{b'} = 0.5$) propagates along the shorter diagonal ($b'$ axis), as illustrated in Fig.~\ref{Fig4}(e).

For the magnetic-structure analysis, we employ a least-squares refinement of the integrated intensities $I_{\rm obs}$ of magnetic Bragg peaks, as in Ref.~\citen{1976_Wil}.
For {\it M} = Al (Ga), we take into account 9 (13) observed peaks and 3 (1) unobserved peaks below $2\theta = 42^{\circ}$ ($44^{\circ}$), where the magnetic reflections are well separated with a few exceptions.
In cases where multiple reflections at close $2\theta$ values appear as a single peak in the ND profile, we take a sum of their calculated intensities ($I_{\rm cal}$) and compare it with $I_{\rm obs}$.

\begin{figure}[t]
\centering
\includegraphics[width=0.93\linewidth]{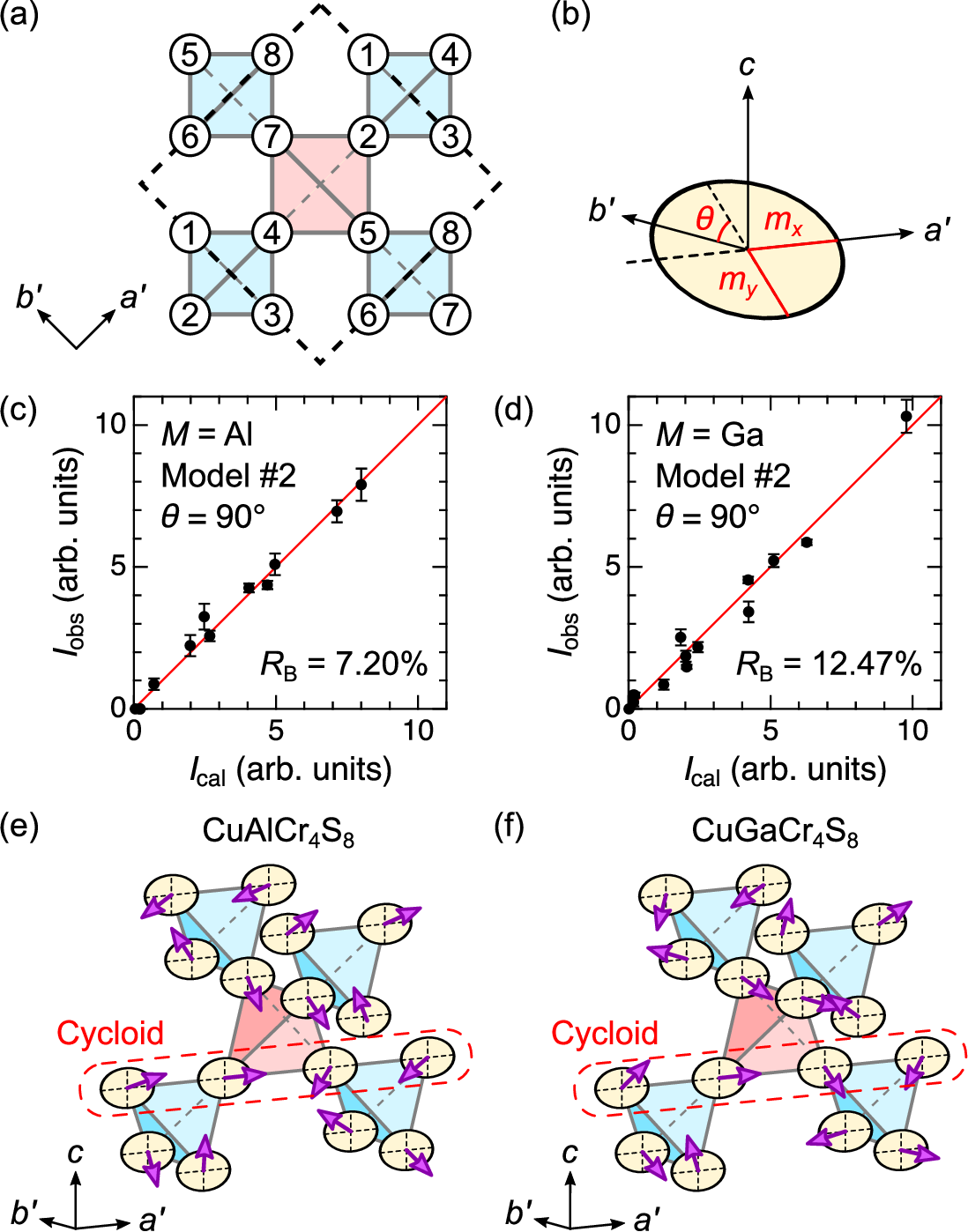}
\caption{(a) Definition of spins 1 $\sim$ 8 in a breathing pyrochlore network projected on the $a'b'$ plane. Odd (even) number spins belong to the Cr1 (Cr2) site [see Fig.~\ref{Fig3}(b)]. Blue and red squares represent small and large tetrahedra, respectively. Black dashed line represents an orthorhombic unit cell. (b) Schematic of an elliptical spiral plane for the tilted cycloidal structure (Model~\#4). $m_{x}$ and $m_{y}$ represent magnetic moments parallel and perpendicular to the $a'$ axis, respectively. $\theta$ is an tilting angle from the $a'b'$ plane. Model~\#2 is a special case of $\theta = 90^{\circ}$. [(c)(d)] $I_{\rm obs}$ versus $I_{\rm cal}$ plot of the magnetic Bragg peaks assuming Models~\#2 for (c) CuAlCr$_{4}$S$_{8}$ and (d) CuGaCr$_{4}$S$_{8}$. [(e)(f)] Schematics of the $a'c$ cycloidal structure obtained by a least-squares refinement for (e) CuAlCr$_{4}$S$_{8}$ and (f) CuGaCr$_{4}$S$_{8}$.}
\vspace{-0.5cm}
\label{Fig5}
\end{figure}

We calculate $I_{\rm cal}$ based on the orthorhombic crystal structure at 4~K revealed by the XRD (Tables~\ref{Tab_Al} and \ref{Tab_Ga}).
In the following, we define 8 distinct spins within the orthorhombic unit cell as spins 1 $\sim$ 8, as depicted in Fig.~\ref{Fig5}(a), where the breathing pyrochlore network is projected on the $a'b'$ plane.
We assume a coplanar structure with a single spiral plane expressed as
\begin{equation}
\begin{split}
\label{Eq_spiral}
{\mathbf m}({\mathbf r}_{{\mathbf l},d})={\mathbf m_{x}}\cos({\mathbf Q} \cdot {\mathbf r}_{{\mathbf l},d}+\phi_{d})+{\mathbf m_{y}}\sin({\mathbf Q} \cdot {\mathbf r}_{{\mathbf l},d}+\phi_{d}),
\end{split}
\end{equation}
where ${\mathbf r}_{{\mathbf l},d}$ and $\phi_{d}$ are the position vector and the phase, respectively, for the spin $d$ ($d$ = 1 $\sim$ 8).
Given that two tetrahedra, consist of spins 1 $\sim$ 4 and 5 $\sim$ 8, respectively, are of the same size, we add a constraint $\delta_{n, 4} = \delta_{n+4, 8}~(n = 1, 2, 3)$, where $\delta_{d, d'} \equiv \phi_{d} - \phi_{d'}$ is defined as the phase difference between spins $d$ and $d'$.
Here, we consider the case of $\delta_{n, n+4} = 0$~(n = 1, 2, 3, 4), resulting in three independent parameters; $\delta_{1, 4}$, $\delta_{2, 4}$, and $\delta_{3, 4}$.
Note that another constraint, $\delta_{n, n+4} = \pi$, can describe the identical magnetic structure obtained for $\delta_{n, n+4} = 0$ because of $q_{b'} = 0.5$.

We consider three types of high-symmetry spiral planes, parallel to the $a'b'$ plane (Model~\#1), the $a'c$ plane (Model~\#2), and the $b'c$ plane (Model~\#3).
Since the incommensurate magnetic modulation propagates along the $a'$ axis, Models~\#1 and \#2 correspond to cycloidal structures, while Model~\#3 corresponds to the proper-screw type.
An ellipticity is introduced so as to match the lattice $Imm2$ symmetry for each model, leading to two degrees of freedom regarding the moment size, $m_{x}$ and $m_{y}$, common to all the spins.
Table~\ref{Tab_RB} summarizes $R_{\rm B}$ values of the final solution obtained by refining five parameters, namely $\delta_{1, 4}$, $\delta_{2, 4}$, $\delta_{3, 4}$, $m_{x}$, and $m_{y}$, for each model.
For both the compounds, Model~\#2 yields the most satisfactory fit, as shown in the $I_{\rm obs}$ versus $I_{\rm cal}$ plots in Figs.~\ref{Fig5}(c) and \ref{Fig5}(d).
On the other hand, Model~\#3 yields a poor fit, suggesting that the ground state is a cycloidal structure but not a proper-screw type one.
 
\begin{table}[t]
\centering
\renewcommand{\arraystretch}{1.2}
\caption{Intensity reliability factors $R_{\rm B}$ of a least-squares fit on the integrated intensity $I_{\rm obs}$ of magnetic Bragg peaks for Cu{\it M}Cr$_{4}$S$_{8}$ ({\it M} = Al, Ga), assuming four types of elliptical helical magnetic structures (see the text for details).}
\begin{tabular}{lrr} \hline
~~~~ & ~~CuAlCr$_{4}$S$_{8}$~~ & ~~CuGaCr$_{4}$S$_{8}$~~ \\ \hline
~Model \#1 ($a'b'$ cycolid)~ & ~11.60\%~~~~ & ~14.72\%~~~~ \\
~Model \#2 ($a'c$ cycolid)~ & ~7.20\%~~~~ & ~12.47\%~~~~ \\
~Model \#3 (proper screw)~ & ~28.83\%~~~~ & ~30.35\%~~~~ \\
~Model \#4 (tilted cycolid)~ & ~6.92\%~~~~ & ~12.18\%~~~~ \\ \hline
\end{tabular}
\vspace{-0.5cm}
\label{Tab_RB}
\end{table}

\begin{table}[t]
\centering
\renewcommand{\arraystretch}{1.2}
\caption{Magnetic structure parameters obtained by a least-squares refinement assuming the $a'c$ cycloidal (Model~\#2) and tilted cycloidal (Model~\#4) structures for Cu{\it M}Cr$_{4}$S$_{8}$ ({\it M} = Al, Ga). For the definition of each parameter, see the text and Fig.~\ref{Fig5}(b).}
\begin{tabular}{cccccccc} \hline
 & \hspace{-0.3cm}Model & ~\hspace{-0.35cm}$\delta_{1, 4}$~ & ~\hspace{-0.45cm}$\delta_{2, 4}$~ & ~\hspace{-0.5cm}$\delta_{3, 4}$~ & ~\hspace{-0.5cm}$m_{x}$~ & ~\hspace{-0.6cm}$m_{y}$~ & ~\hspace{-0.55cm}$\theta$ \\ \hline
\hspace{-0.15cm}{\it M} = Al & \hspace{-0.1cm}\parbox{0.5cm}{\vspace{+0.1cm}{\#2} \\ {\#4} \vspace{+0.1cm}} & \hspace{-0.25cm}\parbox{0.8cm}{\vspace{+0.1cm}{$74(2)^{\circ}$} \\ {$79(2)^{\circ}$} \vspace{+0.1cm}} & \hspace{-0.3cm}\parbox{0.8cm}{\vspace{+0.1cm}{$53(2)^{\circ}$} \\ {$53(2)^{\circ}$} \vspace{+0.1cm}} & \hspace{-0.3cm}\parbox{1.0cm}{\vspace{+0.1cm}{$350(2)^{\circ}$} \\ {$356(2)^{\circ}$} \vspace{+0.1cm}} & \hspace{-0.3cm}\parbox{1.4cm}{\vspace{+0.15cm}{$2.79(2)~\mu_{\rm B}$} \\ {$2.73(2)~\mu_{\rm B}$} \vspace{+0.1cm}} & \hspace{-0.4cm}\parbox{1.4cm}{\vspace{+0.15cm}{$1.39(4)~\mu_{\rm B}$} \\ {$1.53(5)~\mu_{\rm B}$} \vspace{+0.1cm}} & \hspace{-0.4cm}\parbox{0.8cm}{\vspace{+0.1cm}{$90^{\circ}$} \\ {$65(7)^{\circ}$} \vspace{+0.1cm}}\hspace{-0.2cm} \\ \hline
\hspace{-0.15cm}{\it M} = Ga & \hspace{-0.1cm}\parbox{0.5cm}{\vspace{+0.1cm}{\#2} \\ {\#4} \vspace{+0.1cm}} & \hspace{-0.25cm}\parbox{0.8cm}{\vspace{+0.1cm}{$40(2)^{\circ}$} \\ {$47(2)^{\circ}$} \vspace{+0.1cm}} & \hspace{-0.3cm}\parbox{0.8cm}{\vspace{+0.1cm}{$10(2)^{\circ}$} \\ {$10(3)^{\circ}$} \vspace{+0.1cm}} & \hspace{-0.3cm}\parbox{1.0cm}{\vspace{+0.1cm}{$330(2)^{\circ}$} \\ {$338(2)^{\circ}$} \vspace{+0.1cm}} & \hspace{-0.3cm}\parbox{1.4cm}{\vspace{+0.15cm}{$3.09(2)~\mu_{\rm B}$} \\ {$3.02(2)~\mu_{\rm B}$} \vspace{+0.1cm}} & \hspace{-0.4cm}\parbox{1.4cm}{\vspace{+0.15cm}{$1.63(5)~\mu_{\rm B}$} \\ {$1.82(4)~\mu_{\rm B}$} \vspace{+0.1cm}} & \hspace{-0.4cm}\parbox{0.8cm}{\vspace{+0.1cm}{$90^{\circ}$} \\ {$64(5)^{\circ}$} \vspace{+0.1cm}}\hspace{-0.2cm} \\ \hline
\end{tabular}
\vspace{-0.5cm}
\label{Tab_refine}
\end{table}

Furthermore, we consider a tilting angle $\theta$ of the spiral plane from the $a'b'$ plane toward the $a'c$ plane, as shown in Fig.~\ref{Fig5}(b).
We call this ``tilted cycloidal state'' (Model~\#4), which interpolates Model~\#1 ($\theta = 0^{\circ}$) and Model~\#2 ($\theta = 90^{\circ}$).
As shown in Table~\ref{Tab_RB}, the additional introduction of $\theta$ results in a smaller $R_{\rm B}$, although the reduction of $R_{\rm B}$ from Model \#2 is less than 0.3\%.
Table~\ref{Tab_refine} summarizes six parameter sets of the final solution assuming Models~\#2 and \#4 for each compound.
Here, $m_{x}$ ($m_{y}$) is the moment size parallel (perpendicular) to the $a'$ axis.
$\delta_{1, 4}$ $\sim$ $\delta_{3, 4}$ converge to almost the same values regardless of the model.
We hence conclude that the $a'c$ cycloidal state is the most probable structure.
It is necessary to verify whether the magnetic moment along the $b'$ component exists in the future study on single crystals.
Figures~\ref{Fig5}(e) and \ref{Fig5}(f) schematically illustrate the magnetic structure of the $a'c$ cycloidal state for CuAlCr$_{4}$S$_{8}$ and CuGaCr$_{4}$S$_{8}$, respectively.
The spiral plane is found to be elliptically elongated along the $a'$ axis and contracted along the $c$ axis for both the compounds, i.e., $m_{x} > m_{y}$.

\section{\label{Sec4}Discussion}

\begin{figure}[t]
\centering
\includegraphics[width=0.93\linewidth]{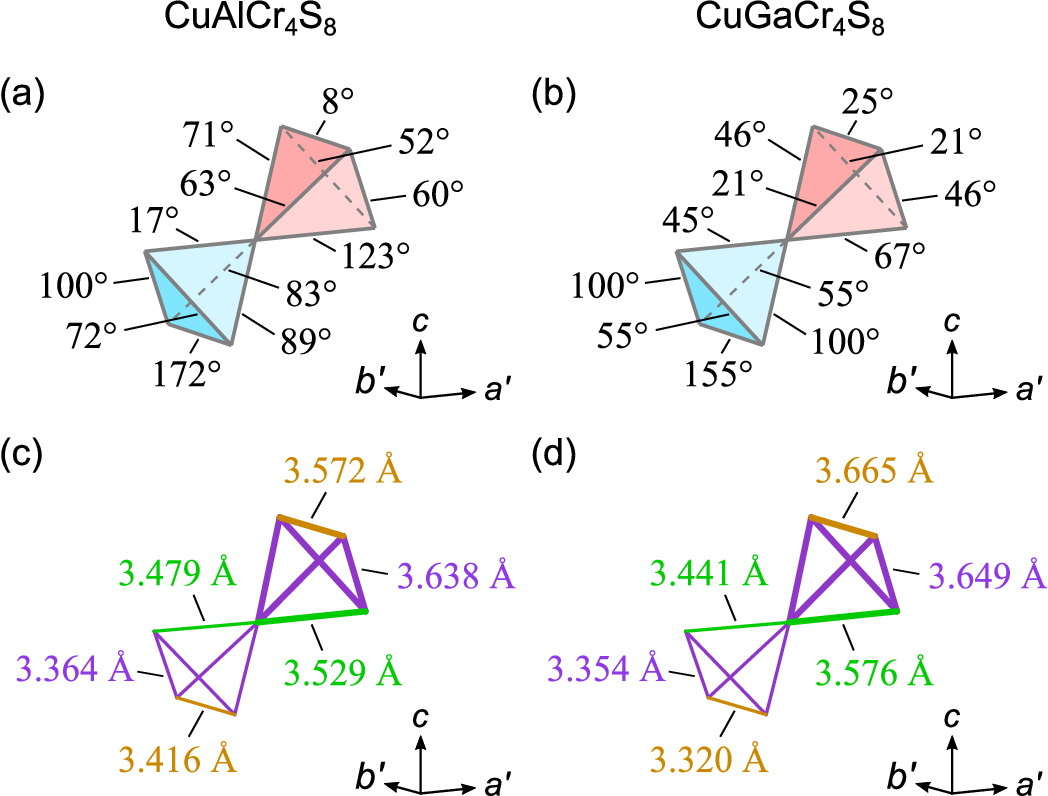}
\caption{[(a)(b)] Relative angles of NN spins for (a) CuAlCr$_{4}$S$_{8}$ and (b) CuGaCr$_{4}$S$_{8}$. Blue and red tetrahedra correspond to small and large tetrahedra, respectively. [(c)(d)] NN Cr--Cr bond lengths at 4~K revealed by the powder XRD experiments for (c) CuAlCr$_{4}$S$_{8}$ and (d) CuGaCr$_{4}$S$_{8}$. Bonds of the same length are drawn in the same color (purple) and thickness.}
\vspace{-0.5cm}
\label{Fig6}
\end{figure}

We have revealed that Cu{\it M}Cr$_{4}$S$_{8}$ ({\it M} = Al, Ga) undergoes a magnetic transition to cycloidal order accompanied by rhombic lattice distortion at low temperatures, where an incommensurate magnetic modulation propagates along the elongated $\langle 110\rangle$ axis of the original cubic cell.
Let us now delve into the detailed discussion of the magnetic structure [Figs.~\ref{Fig5}(e) and \ref{Fig5}(f)].
In Figs.~\ref{Fig6}(a) and \ref{Fig6}(b), we present all the relative angles between NN spins for {\it M} = Al and Ga, respectively.
The corresponding Cr--Cr bond lengths at 4~K, revealed by the crystal structure analysis assuming the $Imm2$ space group [Figs.~\ref{Fig3}(g) and \ref{Fig3}(k)], are shown in Figs.~\ref{Fig6}(c) and \ref{Fig6}(d) for {\it M} = Al and Ga, respectively.

First, it is evident that the spin configuration in the small tetrahedron is different from that in the large tetrahedron for both the compounds, reflecting the breathing nature of the pyrochlore lattice.
For CuGaCr$_{4}$S$_{8}$, the averaged value of ${\mathbf S}_{i} \cdot {\mathbf S}_{j}$ for all the NN spin pairs, $\langle {\mathbf S}_{i} \cdot {\mathbf S}_{j} \rangle$, is 0.10 in the small tetrahedron and 0.75 in the large tetrahedron, assuming $|{\mathbf S}_{i}| = 1$ (Note that $-1/3 \leq \langle {\mathbf S}_{i} \cdot {\mathbf S}_{j} \rangle \leq 1$ for a tetrahedron).
DFT calculations predict $J = 9.8$~K (AFM) and $J'=-11.4$~K (FM) at room temperature \cite{2023_Gen}.
Considering that the volume ratio of the larger tetrahedron to the smaller one changes little across $T_{\rm N}$ [Fig.~\ref{Fig3}(k)], the predominantly FM character of the spin configuration in the large tetrahedron should be attributed to the strong FM coupling $J'$ below $T_{\rm N}$.
For CuAlCr$_{4}$S$_{8}$, $\langle {\mathbf S}_{i} \cdot {\mathbf S}_{j} \rangle$ is 0.04 and 0.39 in the small and large tetrahedron, respectively, exhibiting a smaller difference compared to CuGaCr$_{4}$S$_{8}$.
This may suggest that $J'$ in CuAlCr$_{4}$S$_{8}$ is either more weakly FM or, conversely, AFM.
It is noteworthy that the breathing bond alternation is suppressed within the $ab$ plane in the low-temperature phase in CuAlCr$_{4}$S$_{8}$, as mentioned in Sec.~\ref{Sec3-3}.
Consequently, the strengths of the exchange interactions may also change significantly across $T_{\rm N}$.

Next, we focus on the relation of the local spin configuration to the local lattice distortion.
For both the compounds, a pair of spins on the $d_{a}$ bond (parallel to $a'$ axis) align rather ferromagnetically, while a pair on the $d_{b}$ bond (parallel to the $b'$ axis) antiferromagnetically within the small tetrahedron [Figs.~\ref{Fig6}(a) and \ref{Fig6}(b)].
In contrast, a pair on the $d'_{a}$ bond (parallel to $a'$ axis) align rather antiferromagnetically, and a pair on the $d'_{b}$ bond (parallel to the $b'$ axis) ferromagnetically within the large tetrahedron.
Notably, the FM bond is always longer than the AFM bond within the $a'b'$ plane in each tetrahedron, i.e., $d_{a}>d_{b}$ and $d'_{a}<d'_{b}$ [Figs.~\ref{Fig6}(c) and \ref{Fig6}(d)].
This trend is consistent with the exchange-striction picture for the Cr spinels, in which the AFM configuration favors a shorter Cr--Cr bond length.
As for the remaining four bonds within each tetrahedron, which should be of the same length under the $Imm2$ symmetry, the relative angles between NN spins are not identical to each other.
This is due to the broken mirror symmetry in the magnetic propagation vector ${\mathbf Q} = (q_{\rm IC}, 0.5, 0)_{\rm o}$.
We infer that the crystal structure below $T_{\rm N}$ may indeed exhibit monoclinic $I2$ symmetry, although the current Rietveld analysis based on powder XRD does not allow for the verification of additional crystallographic symmetry breaking.

Our magnetic-structure analysis also reveals that the averaged moment, $m = \sqrt{[(m_{x})^{2}+(m_{y})^{2}]/2}$, is reduced from the full moment of 3~$\mu_{\rm B}$ per a Cr$^{3+}$ spin: $m \approx 2.20(3)$~$\mu_{\rm B}$ for CuAlCr$_{4}$S$_{8}$ and $m \approx 2.47(3)$~$\mu_{\rm B}$ for CuGaCr$_{4}$S$_{8}$ (Table~\ref{Tab_refine}).
Since the orbital degree of freedom is almost quenched in Cr$^{3+}$, the reduction of the ordered moment would be attributed to quantum fluctuation.
Nevertheless, we note that the ordered moment in Cu{\it M}Cr$_{4}$S$_{8}$ are larger than those in most Cr spinel oxides \cite{2018_Gao, 2009_Ji, 2016_Sah}.
This would be because the introduction of the FM exchange interaction $J'$ instead of the AFM $J$ in the large tetrahedron partially alleviates the magnetic frustration.

From a symmetry point of view, Cu{\it M}Cr$_{4}$S$_{8}$ is expected to exhibit cross-correlation responses among spin, charge, and lattice.
As the temperature decreases, the crystal structure undergoes a nonpolar-to-polar transition associated with a magnetic transition at $T_{\rm N}$.
Consequently, the averaged $z$ coordinate of the Cr$^{3+}$ (S$^{2-}$) ions deviates from $z = 0.5$ (Tables~\ref{Tab_Al} and \ref{Tab_Ga}), facilitating spontaneous electric polarization $P_{c}$ along the $c$ axis.
The shift of the average Cr position is estimated to be approximately 0.02~$\AA$ for both the compounds, so that the emergence of subsantial electric polarization is expected.
Moreover, helimagnetism could also contribute to the magnetoelectric effect below $T_{\rm N}$.
The magnetic-structure analysis indicates the development of a cycloidal structure with the spiral plane parallel to the $a'c$ plane, where an incommensurate magnetic modulation propagates along the $a'$ axis.
The crystallographic polarity along the $c$ axis may result in the Dzyaloshinskii--Moriya interaction \cite{1957_Dzy, 1960_Mor} which stabilizes the $a'c$ cycloidal state \cite{2015_Ruf, 2017_Fuj}.
As an inverse effect, this cycloidal structure can generate uniform electric polarization $P_{c}$ along the $c$ axis due to the spin-current mechanism \cite{2005_Kat}, which has been verified in various compounds \cite{2005_Law, 2006_Ari, 2006_Yam, 2006_Kim, 2007_Par}.
The short period of the magnetic modulation, likely arising from the magnetic frustration, should be advantageous for enhancing $P_{c}$.
The combination of the exchange-striction and the spin-current mechanism may bring about magnetoelectric effects in Cu{\it M}Cr$_{4}$S$_{8}$.
Indeed, anomalous dielectric responses have been observed in polycrystalline CuGaCr$_{4}$S$_{8}$ samples associated with magnetic transitions at low temperatures and with the application of a magnetic field \cite{2023_Gen}.
Other intriguing phenomena such as the electric-field control of magnetic domains and the piezoelectric effect may also be possible, although their verification on single crystals remains a future endeavor.

\section{\label{Sec5}Summary}

In summary, we have investigated the crystal and magnetic structures of the breathing pyrochlore magnets Cu{\it M}Cr$_{4}$S$_{8}$ ({\it M} = Al, Ga) at low temperatures through the powder XRD and ND experiments, unveiling the emergence of a qualitatively similar ground state.
At the magnetic transition temperature, Cu{\it M}Cr$_{4}$S$_{8}$ undergoes a structural transition from cubic $F{\overline 4}3m$ to orthorhombic $Imm2$ symmetry, characterized by polar space group.
The magnetic structure manifests an incommensurate cycloidal state with a magnetic propagation vector of ${\mathbf Q} = (q_{\rm IC}, 0.5, 0)_{\rm o}$ in the orthorhombic setting, where $q_{\rm IC} \approx 0.39$ for {\it M} = Al and $q_{\rm IC} \approx 0.31$ for {\it M} = Ga.
The larger deviation of $q_{\rm IC}$ from 0.5 for CuGaCr$_{4}$S$_{8}$ aligns with the observed larger rhombic distortion in terms of the exchange-striction mechanism.
We also confirm a strong correlation between the local spin configuration and the local lattice distortion, underscoring the significance of spin-lattice coupling in determining the ground state.
Remarkably, the low-temperature phase of Cu{\it M}Cr$_{4}$S$_{8}$ is free from a phase separation often seen in other chromium spinels \cite{2007_Mat, 2016_Sah, 2015_Nil, 2009_Yok}, suggesting that the identified spin-lattice-coupled helical magnetic order is robust against minor perturbations in the Cu{\it M}Cr$_{4}$S$_{8}$ system.
We propose that Cu{\it M}Cr$_{4}$S$_{8}$ serves as a promising platform for investigating magnetoelectric effects arising from the emergent crystallographic polarity as well as helimagnetism.

\section*{Acknowledgments}

This work was financially supported by the JSPS KAKENHI Grants-In-Aid for Scientific Research (No.~23K13068).
The authors thank A. Ikeda for generously allowing the use of optical sensing instrument (Hyperion si155, LUNA) for thermal expansion measurements.
The authors thank K. Adachi and D. Hashizume for the support of the in-house XRD experiments at Materials Characterization Support Team, RIKEN CEMS.
The powder ND experiments at JRR-3 were carried out along the proposal No.~23401 and partly supported by ISSP of the University of Tokyo.
The authors thank S. Kitou for fruitful discussions.

\end{document}